\documentclass[twocolumn]{aastex61}
\usepackage{amsmath,amstext}
\usepackage{color}
\usepackage[toc,page]{appendix}
\shorttitle{Haumea's Shape and Composition}
\shortauthors{Dunham, E.~T. {\it et al.}}

\begin{document}

\title{Haumea's shape, composition, and internal structure}

\author{Dunham, E.~T}
\affiliation{School of Earth and Space Exploration, Arizona State University}
\author{Desch, S.~J.}
\affiliation{School of Earth and Space Exploration, Arizona State University}
\author{Probst, L.}
\affiliation{San Francisco University High School}
\begin{abstract}

We have calculated the figure of equilibrium of a rapidly rotating, differentiated body to determine the shape, structure, and composition of the dwarf planet Haumea.  Previous studies of Haumea's light curve have suggested Haumea is a uniform triaxial ellipsoid consistent with a Jacobi ellipsoid with axes $\approx 960 \times 774 \times 513$ km, and bulk density $\approx 2600 \, {\rm kg} \, {\rm m}^{-3}$.  In contrast, observations of a recent stellar occultation by Haumea indicate its axes are $\approx 1161 \times 852 \times 523$ km and its bulk density $\approx 1885 \, {\rm kg} \, {\rm m}^{-3}$; these results suggest that Haumea cannot be a fluid in hydrostatic equilibrium and must be partially supported by interparticle forces. We have written a code to reconcile these contradictory results and to determine if Haumea is in fact a fluid in hydrostatic equilibrium. The code calculates the equilibrium shape, density, and ice crust thickness of a differentiated Haumea after imposing (semi-) axes lengths $a$ and $b$. We find Haumea is consistent with a differentiated triaxial ellipsoid fluid in hydrostatic equilibrium with axes of best fit $a$ = 1050 km, $b$ = 840 km, and $c$ = 537 km. This solution for Haumea has  $\rho_{\rm avg} = 2018 \, {\rm kg} \, {\rm m}^{-3}$, $\rho_{\rm core} = 2680 \, {\rm kg} \, {\rm m}^{-3}$, and core axes $a_{\rm c} = 883$ km, $b_{\rm c} = 723$ km, and $c_{\rm c}=470$ km, which equates to an ice mantle comprising $\sim 17\%$ of Haumea's volume and ranging from 67 to 167 km in thickness. The thick ice crust we infer allows for Haumea's  collisional family to represent only a small fraction of Haumea's pre-collisional ice crust. For a wide range of parameters, the core density we calculate for Haumea suggests that today the core is composed of hydrated silicates and likely underwent serpentinization in the past. 

\end{abstract}

\section{Introduction}

The Kuiper Belt Object (KBO) and dwarf planet Haumea is one of the most intriguing and puzzling objects in the outer Solar System. 
Haumea orbits beyond Pluto, with a semi-major axis of 43.2 AU, and is currently near its aphelion distance $\approx 51.5$ AU, but is relatively bright at magnitude $V = 17.3$, due to its large size and icy surface.
Haumea's mean radius is estimated to be $\approx 720$ km \citep{Lockwood2014} to $\approx 795$ km \citep{Ortiz2017}, and its reflectance spectra indicate that Haumea's surface is uniformly covered by close to 100\% water ice 
\citep{Trujillo2007, Pinilla-Alonso2009}. 
Haumea is the third-brightest KBO, after the dwarf planets Pluto and Makemake \citep{Brown2006}.
Haumea has two small satellites, Hi'iaka and Namaka, which enable a determination of its mass, $M_{\rm H} = 4.006 \times 10^{21} \, {\rm kg}$ \citep{Ragozzine2009};
it is the third or fourth most massive known KBO (after Pluto, Eris, and possibly Makemake). Despite its large size, it is a rapid rotator; from its light curve Haumea's rotation rate is found to be $3.91531 \pm 0.00005$ hours \citep{Lellouch2010}. This means Haumea is the fastest-rotating KBO \citep{Sheppard2002}, and is in fact the fastest-rotating large ($> 100$ km) object in the Solar System \citep{Rabinowitz2006}. Haumea is also associated with a collisional family \citep{Brown2007} and is known to have a ring \citep{Ortiz2017}. Based on its rapid rotation and its collisional family, Haumea is inferred to have suffered a large
collision \citep{Brown2007}, $> 3$ Gyr ago, based on the orbital dispersion of the family members
\citep{Volk2012}. 

Haumea is larger than other dwarf planets such as Ceres (radius 473 km), or satellites such as Dione (radius 561 km)
or Ariel (radius 579 km), all of which are nominally round.
Despite this, Haumea exhibits a reflectance light curve with a very large peak-to-trough amplitude, $\Delta m \approx 0.28$ in 2005 \citep{Rabinowitz2006}, $\Delta m = 0.29$ in 2007 \citep{Lacerda2008}, and $\Delta m = 0.32$ in 2009 \citep{Lockwood2014}. 
Since Haumea's surface is spectrally uniform, such an extreme change in brightness can only be attributed to a difference in the area presented to the observer. Haumea has been modeled as a triaxial ellipsoid with axes $a > b > c$, the $c$-axis being aligned with the rotation axis. In that case, the change in brightness from peak to trough would be given by
\begin{equation}
\Delta m = 2.5 \, \left[ \log_{10} \left( \frac{ a }{ b } \right) - \log_{10} \left( \frac{ r_1 }{ r_2 } \right) \right],
\end{equation} 
where 
\begin{equation}
r_1 = \left( a^2 \cos^2 \phi + c^2 \sin^2 \phi \right)^{1/2}
\end{equation} 
and 
\begin{equation}
r_2 = \left( b^2 \cos^2 \phi + c^2 \sin^2 \phi \right)^{1/2},
\end{equation} 
$\phi$ being the angle between the rotation axis and the line of sight \citep{Binzel1989AsteroidsII}. 
If $\phi = 0^{\circ}$, then $\Delta m = 0$, because the same $a \times b$ ellipse would be presented to the observer.
Instead, $\Delta m$ is maximized when $\phi = 90^{\circ}$, because then the ellipse presented to the observer would
vary between $a \times c$ and $b \times c$.
In that case, $\Delta m = 2.5 \, \log_{10} (a / b)$ and the axis ratio is related directly to $\Delta m$. 
Assuming $\phi = 90^{\circ}$ in 2009, when $\Delta m = 0.32$, one would derive $b / a = 0.75$. 
Taking into account the scattering properties of an icy surface, \citet{Lockwood2014} refined this to 
$b / a = 0.80 \pm 0.01$. 
Thus, Haumea is distinctly non-spherical, and is not even axisymmetric.

Haumea appears to be unique among large Solar System objects in having such a distinctly non-axisymmetric, triaxial 
ellipsoid shape.
Based on its rapid rotation (angular velocity $\omega = 4.457 \times 10^{-4} \, {\rm s}^{-1}$), Haumea is inferred to 
have assumed a particular shape known as a Jacobi ellipsoid. This is a class of equilibrium shapes assumed by (shearless) fluids 
in hydrostatic equilibrium when they rotate faster than a certain threshold (Chandrasekhar 1969, 1987). 
For a body with angular velocity $\omega$ and uniform density $\rho$, the axis ratios $b / a$ and $c / a$ of the ellipsoid
are completely determined, and the axis ratios are single-valued functions of $\omega^{2} / (\pi G \rho)$. 
For a Jacobi ellipsoid with $b / a = 0.806$ and Haumea's rotation rate, the density must be
$\rho = 2580 \, {\rm kg} \, {\rm m}^{-3}$, and $c / a = 0.520$.
Assuming a semi-axis of $a = 960$ km then yields $b = 774$ km and then $c = 499$ km, $(4\pi/3) a b c \, \rho$ 
exactly matches Haumea's mass. 
The mean radius of Haumea would be 718 km.
Moreover, the cross-sectional area of Haumea would then imply a surface albedo $p_{\rm V} \approx 0.71 - 0.84$
\citep{Rabinowitz2006,Lacerda2006, Lellouch2010, Lockwood2014},
consistent with the albedo of a water ice surface.

To explain Haumea's icy surface and $\rho = 2580 \, {\rm kg} \, {\rm m}^{-3}$, one would have to assume that the interior of Haumea was close to $2600 \, {\rm kg} \, {\rm m}^{-3}$
in density (an interior of hydrated silicates \citep{Desch2015}) while its surface was a very thin ice layer (\S\ref{sec:2}). This structure implies that Haumea suffered a giant collision in its past that may have stripped its ice mantle. For these reasons, the above axes and axis ratios were strongly favored in the literature. Other groups derived similar axes and bulk densities \citep{Rabinowitz2006, Lacerda2008, Lellouch2010} 

This model was upended by the observations of \citet{Ortiz2017} following the occultation of an $18^{\rm th}$ magnitude star by Haumea in January 2017. The shadow of Haumea traced out an ellipse, as expected for the shadow of a triaxial ellipsoid; but the (semi-)axes of the shadow ellipse were much larger than expected: $b' = 569 \pm 13$ km by $a' = 852 \pm 2$ km. 
\citet{Ortiz2017} used the shadow axes and other assumptions to derive the axes of Haumea to be  $a  = 1161 \pm 30$, $b = 852 \pm 4$, and $c = 513 \pm 16$ km (\ref{app}). 


This new shape causes Haumea to look significantly different than previous models: the mean radius of Haumea is larger at 798 km, the albedo is a smaller $p_{\rm V} \approx 0.51$ (and would require a darkening agent in addition to water ice), and the bulk density a lower $1885 \pm 80 \, {\rm kg} \, {\rm m}^{-3}$.
Moreover, the axis ratio $c / a \approx 0.44$ is significantly lower than previous estimates $\approx 0.52$,
and is inconsistent with a Jacobi ellipsoid or a fluid in hydrostatic equilibrium.
\citet{Ortiz2017} point out the possibility that shear stresses may be supported on Haumea by granular interparticle forces \citep{Holsapple2001}.


In either case, Haumea is likely to have a rocky core surrounded by ice, but no analytical solution exists for the figure of equilibrium of a rapidly rotating, differentiated body. Therefore it is not known whether or not Haumea is a fluid in hydrostatic equilibrium. In this paper we attempt to reconcile the existing data from Haumea's light curve and occultation shadow, with the goal of deriving its true shape and internal structure.
Besides its axes, important quantities to constrain are the ice fraction on Haumea today, and the size, shape, and
density of its core. A central question we can solve using these quantities is whether Haumea is a fluid in hydrostatic equilibrium or demands granular physics to support it against shear stresses. In addition, we can use the core density and ice fraction to constrain the geochemical evolution of Haumea and, by extension, other KBOs,
as well as models of the origin of Haumea's collisional family.

In \S \ref{sec:2}, we examine whether it is possible for Haumea to be a differentiated (rocky core, icy mantle) body with a Jacobi ellipsoid shape. We show that only homogeneous bodies
are consistent with a Jacobi ellipsoid shape. 
In \S \ref{sec:3} we describe a code we have written to calculate the equilibrium figure of a rapidly rotating, differentiated body.
In \S \ref{sec:4} we present the results, showing that entire families of solutions exist that allow Haumea to be a 
differentiated body in hydrostatic equilibrium. 
Some of these solutions appear consistent with observations of Haumea, particularly $\approx 1050 \times 840 \times 537$ km with bulk density $\approx 2018 \, {\rm kg} \, {\rm m}^{-3}$.
In \S \ref{sec:5} we discuss the implications of this solution for Haumea's structure, for the collision that created the collisional family, and for the astrobiological potential of Haumea.  


\section{Can a differentiated Haumea have Jacobi ellipsoid axes?\label{sec:2}}

Observations have suggested that Haumea has axes consistent with a Jacobi ellipsoid: for example,
\citet{Lockwood2014} inferred axes of $a = 960$ km, $b = 770$ km, and $c = 495$ km
(yielding axis ratios $b / a = 0.802$, $c / a = 0.516$) 
and a uniform density $\approx 2614 \, {\rm kg} \, {\rm m}^{-3}$.
These axes are within 1\% of the Jacobi ellipsoid solution: 
a Jacobi ellipsoid with Haumea's mass, rotation rate, and $a = 960$ km, has axes
$b = 774.2$ km and $c = 498.8$ km (yielding axis ratios $b / a = 0.806$ and $c / a = 0.520$),
and uniform density $2580 \, {\rm kg} \, {\rm m}^{-3}$. 
Uniform density is a central assumption of the Jacobi ellipsoid solution, and yet Haumea is manifestly
not uniform in density.
Its reflectance spectra robustly show the existence of a uniform water-ice surface \citep{Trujillo2007,Pinilla-Alonso2009}. The density of this ice, which has structure Ih at 40K, is about $\approx 921 \, {\rm kg} \, {\rm m}^{-3}$ \citep{Desch2009}, much lower than Haumea's mean density. 
Haumea, therefore, is certainly differentiated.
A basic question, then, is whether a differentiated Haumea can be consistent with axes that
match a Jacobi ellipsoid.

To answer this question, we have calculated the gravitational potential of a differentiated Haumea,
as described in \citet{Probst2015}. 
We model Haumea as two nested, aligned triaxial ellipsoids. 
We assume Haumea's outer surface is an ellipsoid with axes $a = 960$ km, $b = 770$ km, and $c = 495$ km,
and we allow its core to have arbitrary density $\rho_{\rm core}$, and arbitrary axis ratios
$p_{\rm c} = b_{\rm c} / a_{\rm c}$ and $q_{\rm c} = c_{\rm c} / a_{\rm c}$.
For a given density $\rho_{\rm core}$ and axis ratios $p_{\rm c}$ and $q_{\rm c}$, the core axis
$a_{\rm c}$ is chosen so that the mass of the core plus the mass of the ice mantle, with density
$\rho_{\rm ice} = 921 \, {\rm kg} \, {\rm m}^{-3}$, 
equal the mass of Haumea.
We then calculate whether the surface and the core-mantle boundary (CMB) are equipotential surfaces.

\begin{figure*}
\centering
\includegraphics[width=6.5in]{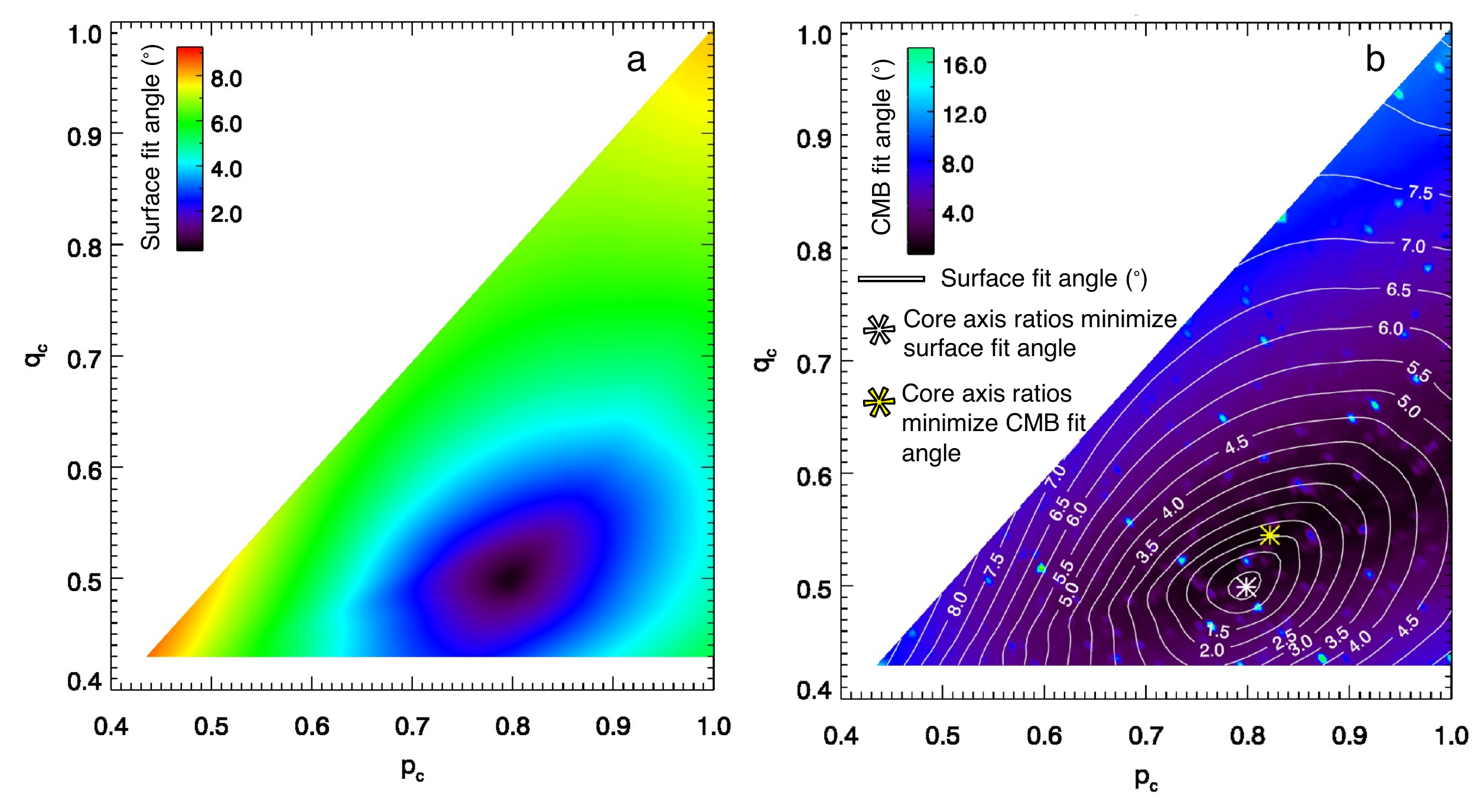}
\caption{The ``fit angle" $\Theta$ on the surface ({\it a}) and the core-mantle boundary ({\it b}), 
as functions of the core axis ratios $p_{\rm c} = b_{\rm c} / a_{\rm c}$ and $q_{\rm c} = c_{\rm c} / a_{\rm c}$
(note that $p_{\rm c} \geq q_{\rm c}$).  A core density $2700 \, {\rm kg} \, {\rm m}^{-3}$ has been assumed.
The core axis ratios that minimize the surface fit angle and are most consistent with equilibrium are 
$p_{\rm c} \approx 0.80$ and $q_{\rm c} \approx 0.51$ (denoted by the white star in {\it b}).
The core axis ratios that minimize the fit angle on the CMB are $p_{\rm c} \approx 0.82$ and 
$q_{\rm c} \approx 0.55$ (denoted by the yellow star in {\it b}).  
On either surface the fit angle is at least a few degrees, and it is not possible
to minimize $\Theta$ on both surfaces with the same core ratios.}
\label{figure:highres}
\end{figure*} 
An equilibrium solution must have the equipotential surfaces coincident with the surface and CMB, or 
else vortical flows will be generated.
In the absence of external and viscous forces and large internal flows, the vorticity $\overrightarrow{\omega}$ 
follows the equation 
\begin{equation}
\frac{D \overrightarrow{\omega} }{D t} = 
\frac{1}{\rho^2} \, \overrightarrow{ \nabla } \rho \times \overrightarrow{ \nabla } P
\end{equation}
where $P$ is the pressure and $\rho$ is the density. 
If the gradients of $\rho$ and $P$ are misaligned by an angle $\Theta$, then vorticity will be generated 
at a rate $\sim (P / \rho) / R^2 (\sin \Theta) $, where $R$ is comparable to the mean radius.
In a time $\tau$, the vortical flows will circulate at rates comparable to the rotation rate, $\omega_{\rm H}$,
where $\tau \sim \omega_{\rm H} \rho R^2 / P \, \left( \sin \Theta \right)^{-1}$. 
Assuming $\omega_{\rm H} = 4.46 \times 10^{-4} \, {\rm s}^{-1}$, $\rho = 921 \, {\rm kg} \, {\rm m}^{-3}$,
$P = 18$ MPa, and $R \sim 725$ km, the timescale $\tau \sim 4 / \sin \Theta$ years.
Even a small mismatch, with $\Theta \sim 1^{\circ}$, would lead to significant vortical flows within
hundreds of years.
Equilibrium solutions demand $\Theta = 0^{\circ}$, i.e., that $\overrightarrow{\nabla} \rho$ and 
$\overrightarrow{\nabla} P$ are parallel. 
In hydrostatic equilibrium, the net force is 
\begin{equation}
\overrightarrow{F} = -\overrightarrow{\nabla} P + \rho \overrightarrow{g}_{\rm eff},
\end{equation}
where $\overrightarrow{g}_{\rm eff}$ measures the acceleration due to gravity as well as centrifugal support due 
to Haumea's rotation. 
If the net force is zero, then it must be the case that $\overrightarrow{\nabla} \rho$ is parallel to 
$\overrightarrow{g}_{\rm eff}$.
In other words, the gradient in the effective gravitational potential must be parallel to the gradient in 
density, and surfaces of density discontinuity must be equipotential surfaces. 

We solve for the effective gravitational potential by discretizing an octant of Haumea on a Cartesian grid with 
60 evenly spaced zones along each axis. 
If a rectangular zone is entirely inside the triaxial ellipsoid defined by the core, its density is 
set to $\rho_{\rm core}$; if it is entirely outside the triaxial ellipsoid defined by the surface, its
density is set to zero; and if it is entirely between these two ellipsoids, its density is set to $\rho_{\rm ice}$.  
For zones straddling the core and ice mantle, or straddling the ice mantle and the exterior, the density is
found using a Monte Carlo method. 
An array of about 100 points on the surface is then generated, by generating a grid of $N_{\theta} \approx 10$ angles 
$\theta$ from 0 to $\pi$ radians, and of $N_{\phi} \approx 10$ angles $\phi$ from 0 to $2\pi$ radians. 
The points on the surface are defined by $x^{*} = a \, \sin \theta \, \cos \phi$, $y^{*} = b \, \sin \theta \, \sin \phi$,
$z^{*} = c \, \cos \theta$. 
At each point we find the vector normal to the surface,
$\overrightarrow{n} = (2 x^{*} / a^2) \hat{e}_{x} + (2 y^{*} / b^2) \hat{e}_{y} + (2 z^{*} / c^2) \hat{e}_{z}$, 
as well as the gravitational acceleration $\overrightarrow{g}$, found by summing the gravitational acceleration vectors
from each zone's contribution.
To this we add an additional contribution due to centrifugal support:
\begin{equation}
\overrightarrow{g}_{\rm eff} = \overrightarrow{g} + \omega^2 x^{*} \, \hat{e}_{x} + \omega^{2} y^{*} \, \hat{e}_{y}.
\end{equation}
Once these vectors are found at each of the surface points defined by $\theta$ and $\phi$, we find the following
quantity by summing over all points:
\begin{equation}
{\cal M} = \frac{1}{ N_{\theta} N_{\phi} } \, \sum_{N_{\theta}} \sum_{N_{\phi}} \, 
 \frac{ \overrightarrow{n} \cdot \overrightarrow{g}_{\rm eff} }
      { \left| \overrightarrow{n} \right| \, \left| \overrightarrow{g}_{\rm eff} \right| } .
\end{equation}
We also define ``fit angle," the mean angular deviation between the surface normal and the effective gravitational field:
\begin{equation}
\Theta = \cos^{-1} \, {\cal M}.
\end{equation}
If the equipotential surfaces are coincident with the surface, then ${\cal M} = 1$ and $\Theta = 0^{\circ}$.
In a similar fashion we define an identical metric ${\cal M}$ on the core-mantle boundary as well. 

In Figure~\ref{figure:highres}, we plot the fit angle on the surface and core-mantle boundary of Haumea,
as a function of the assumed core axis ratios $p_{\rm c}$ and $q_{\rm c}$.
A core density $2700 \, {\rm kg} \, {\rm m}^{-3}$ is assumed.
The core axis ratios that minimize $\Theta$ on the outer surface are $p_{\rm c} = b_{\rm c} / a_{\rm c}$
$\approx 0.80$ and $q_{\rm c} = c_{\rm c} / a_{\rm c}$ $\approx 0.51$. For this combination, 
$\Theta \approx 1^{\circ}$.
Meanwhile, the core axis ratios that minimize $\Theta$ on the core-mantle boundary are 
$p_{\rm c} \approx 0.82$ and $q_{\rm c} \approx 0.55$. For this combination, $\Theta \approx 2^{\circ}$.
Significantly, the parameters that minimize $\Theta$ on the outer surface are not those that minimize
$\Theta$ on the core-mantle boundary. 
Given the change in $\Theta$ with changes in $p_{\rm c}$ and $q_{\rm c}$, either $\Theta$ on the surface
or core-mantle boundary must be at least a few degrees. Figure \ref{figure:highres} shows numerous patches of higher-than-expected angles. These patches are caused by the random nature in which the grid cells straddling the CMB are populated. The coarseness of the grid leads to the code creating a bumpiness in the CMB surface which then upwardly skews the calculation of the average fit angle in places. This is because where the surface is bumpy, the surface normal vector can be significantly different from the gravitational acceleration vector. The patches become more numerous, but smaller in magnitude, with increasing numerical resolution.

We have repeated the analysis for other core densities of $3000 \, {\rm kg} \, {\rm m}^{-3}$ and 
$3300 \, {\rm kg} \, {\rm m}^{-3}$. In those cases the discrepancy between what parameters $p_{\rm c}$ and
$q_{\rm c}$ minimize $\Theta$ on the surface vs. what parameters minimize it on the CMB grows even larger.
If Haumea has axes $a = 960$ km, $b = 770$ km, and $c = 495$ km, and is divided into a rocky core and icy mantle, 
the only way to maintain hydrostatic equilibrium on the surface and core-mantle boundary is for the core
density to be as close as possible to the inferred bulk density of Haumea, $\approx 2600 \, {\rm kg} \, {\rm m}^{-3}$,
and for the core and surface axis ratios to converge. 
The solution is naturally driven to one of uniform density.
In this case, the core comprises over 96\% of the mass of Haumea, and the ice thickness is $< 10$ km on the 
$a$ and $b$ axes, and $< 5$ km on the $c$ axis.  Even for this case, though, the effective equipotential surfaces fail to coincide with the surface or core-mantle boundary, by several degrees. 
This suggests that the only way for Haumea to have axes consistent with a Jacobi ellipsoid is for it to have essentially no ice mantle, less than a few km thick. 

These investigations reveal two facts.
First, a Haumea divided into a rocky core and icy mantle cannot have axes equal to those of a Jacobi ellipsoid.
This lends some support to the finding by \citet{Ortiz2017} that Haumea's axes deviate significantly
from a Jacobi ellipsoid's.
Second, if Haumea cannot conform to a Jacobi ellipsoid, then it is not possible to use analytical formulas
to describe its shape, and a more powerful technique must be used to derive its internal structure.


\section{Methods\label{sec:3}}

To calculate the shape of a rapidly rotating Haumea with a rocky core and icy mantle, we 
have written a code, named {\tt kyushu}, that calculates the internal structure and figure
of equilibrium of a differentiated body undergoing rapid, uniform rotation.
Our algorithm is adapted from that of Hachisu (1986a,b; hereafter H86a,H86b), for calculating 
the structure of stars orbiting each other in binary systems, as follows.

The \citet{Hachisu1986a, Hachisu1986} algorithm relies on using a governing equation derived from the Bernoulli 
equation,  at each location on a three-dimensional grid:
\begin{equation}
\int \rho^{-1} \, dP + \Phi - \int \Omega^2 r_{\perp} \, dr_{\perp} = C, 
\end{equation}
where the first term is the enthalpy, $H$, the second term is the gravitational potential
energy, the third term is the rotational energy, and $C$ is a constant.
Here $r_{\perp}$ is the distance from the rotation axis. 

The grid is defined in spherical polar coordinates, the variables being distance from the origin, $r$,
the cosine of the polar angle, $\mu$, and the azimuthal angle, $\phi$.
A discretized grid of $r_i$, with $i = 1, 2, ... N_r$ is defined, with $r$ uniformly spaced between 
$r_1 = 0$ and $r_{Nr} = R$.
Likewise, a discretized grid of $\mu_j$, with $j = 1, 2, ... N_{\mu}$ is defined, with $\mu$ uniformly
spaced between $\mu_1 = 0$ and $\mu_{N\phi} = 1$, and a discretized grid of $\phi_{k}$, with 
$k = 1, 2, ... N_{\phi}$, is defined, with $\phi$ uniformly spaced between $\phi_1 = 0$ and 
$\phi_{N\phi} = \pi / 2$. 
Symmetries across the equatorial plane and $n=2$ symmetry about the polar axis are assumed.
Quantities in the above equation, including density $\rho_{ijk}$, gravitational potential 
$\Phi_{ijk}$, etc., are defined on the intersections of grid lines. 
Typical values in our calculation are $R = 1300 \, {\rm km}$, $N_{r} = 391$, $N_{\mu} = 33$,
and $N_{\phi} = 33$, meaning that quantities are calculated at $33 \times 33 \times 391 = 425,799$ locations.

The enthalpy term can be calculated if the density structure and equation of state are provided. 
For example, if $P = K \rho^{\gamma}$ (as for an adiabatic gas), then 
$H = (\gamma) / (\gamma-1) P / \rho$, and is immediately known as a function of the local pressure 
and density. For planetary materials such as olivine, clays, or water ice, it would be more appropriate to use a Vinet \citep{Vinet1987} or Birch-Murnaghan \citep{Birch2947} equation of state, with the bulk modulus and the pressure derivative of the bulk modulus specified. This equation of state can then be integrated to yield the enthalpy. The recent paper by \cite{Price2019} provides formulas for this. For our purposes, we neglect the self-compression of the planetary materials inside Haumea. The bulk moduli of planetary materials are typically 10s of GPa, while the maximum pressure inside Haumea is $<$ 0.4 GPa, so self-compression can be ignored. For ease of calculation we therefore assume uniform densities in the ice mantle and in the rocky core, and compute the enthalpy accordingly.

The gravitational potential term is found by numerical integration of an expansion of the
gravitational potential in spherical harmonics, using equations 2, 3, 33, 34, 35 and 36 of H86b,
using $n = 2$ symmetry, and typically $N_l = 16$ terms in the expansion. 
The rotational term is defined to be $\Omega^2 \Psi$, where $\Psi = -r_{\perp}^2 / 2$, 
$r_{\perp}$ again being the distance from the axis.
The terms $H$, $\Phi$ and $\Omega^2 \, \Psi$ all spatially vary, but their sum is a constant $C$
at all locations. 
The Hachisu algorithm exploits this fact by fixing two spatial points ``A" and ``B" to be on the 
boundary of the body.
Point ``A" lies at $r = r_A$, $\mu = 0$ (in the equatorial plane), $\phi = 0$,
or $x = r \, (1 - \mu^2)^{1/2} \, \cos \phi = r_A$, $y = r \, (1 - \mu^2)^{1/2} \, \sin \phi = 0$, 
$z = r \, \mu = 0$.
Point ``B" lies at $r = r_B$, $\mu = 0$ (in the equatorial plane), $\phi = \pi / 2$,
or $x = 0$, $y = r_B$, $z = 0$.
For a triaxial ellipsoid $(x / a)^2 + (y / b)^2 + (z / c)^2 = 1$, points $A$ and $B$
refer to the long and intermediate axes of the body in the equatorial plane, with 
$r_A = a$ and $r_B = b$.
At these two locations, $H = 0$, and the Hachisu algorithm then solves for the only two values of $C$
and $\Omega$ that allow $H = 0$ at both these boundary points.
With $C$ and $\Omega$ defined, $H$ is found at all locations, and the enthalpy integral
$H = \int \rho^{-1} \, dP$ is inverted to find the density $\rho$ at each location. 
Locations with $H < 0$ are assigned zero density. 
The farthest point with non-zero density along the $z$ axis can be equated with $c$ of a triaxial
ellipsoid, although of course the shape need not necessarily be a triaxial ellipsoid. 
After adjusting the density everywhere, the code then recalculates the gravitational potential and 
performs the same integrations, solving iteratively until the values of $\Omega$ and the densities 
$\rho$ at all locations converge.
Because the densities and the volume of the body are changed with each iteration, the mass of the
object is a varying output of the model. 

We apply the H86a,b algorithms as part of a larger iterative procedure that introduces two new 
variables to an equation of state: $P_{\rm cmb}$ and $\rho_{\rm core}$.
We assume that at locations within the body with pressures $P < P_{\rm cmb}$, 
$\rho = \rho_{\rm ice} \equiv 921 \, {\rm kg} \, {\rm m}^{-3}$.
At higher pressures $P > P_{\rm cmb}$, we assume $\rho = \rho_{\rm core}$.
This divides the body into a core and an icy mantle, each with distinct densities, with the 
pressure equal to a uniform value $P_{\rm cmb}$ everywhere on the core-mantle boundary. For $P > P_{\rm cmb}$, 
$H = P_{\rm cmb} / \rho_{\rm ice} + (P - P_{\rm cmb}) / \rho_{\rm core}$.
This form of the equation of state ignores self-compression, as is appropriate in bodies of Haumea's size made of materials like water ice, olivine, or hydrated silicates. The bulk modulus of water ice is 9.2 GPa \citep{Shaw1986}, far higher than the likely pressures in the ice shell, $<$ 20 MPa (section \ref{sec:4}). Likewise, the bulk moduli of olivine is 126 GPa \citep{Nunez-Valdez2013}, and that of the hydrated silicate clay antigorite is 65 GPa \citep{Capitani2012}, far higher than the maximum pressure inside Haumea ($<$ 300 MPa). We therefore expect self-compression to change the densities by $<$1\%, and we are justified in assuming uniform densities. This equation of state allows a very simple inversion to find the density:
for $H < P_{\rm cmb} / \rho_{\rm ice}$, the density is simply $\rho_{\rm ice}$, and for 
$H > P_{\rm cmb} / \rho_{\rm ice}$, $\rho = \rho_{\rm core}$.

With this definition, we iterate as follows to find $P_{\rm cmb}$ and $\rho_{\rm core}$.
In each application of the Hachisu algorithms, we initialize with a density distribution 
with $\rho = 0$ for $(x / a)^2 + (y / b)^2 + (z / c)^2 > 1$, 
$\rho = \rho_{\rm ice}$ for $(x / a)^2 + (y / b)^2 + (z / c)^2 < 1$, 
but $\rho = \rho_{\rm core}$ for $(x / a)^2 + (y / b)^2 + (z / c)^2 < \xi^{2}$.
That is, we define Haumea's surface to be a triaxial ellipsoid, and its core to be 
a similar triaxial ellipsoid with aligned axes, smaller in size by a factor of $\xi$.
The value of $\xi$ is chosen so that the total mass of the configuration equals $M_{\rm H}$,
the mass of Haumea:
\begin{equation}
\xi = \left[ \frac{ 3 M_{\rm H} / (4\pi a b c) - \rho_{\rm ice} }{ \rho_{\rm rock} - \rho_{\rm ice} } \right]^{1/3}.
\end{equation}
The value of $c$ is unknown and an output of the code, so we initialize our configuration with $c = b$.
We apply the Hachisu algorithms several times. 
First we define $P_{\rm cmb} \approx 0 \, {\rm MPa}$ and define $\rho_{\rm core} = 2700 \, {\rm kg} \, {\rm m}^{-3}$, 
apply the Hachisu algorithms, and calculate the mass of the body, $M$. 
Because this will not match Haumea's mass, $M_{\rm H}$, we multiply $\rho_{\rm core}$ by a factor
$M_{\rm H} / M$, and reapply the Hachisu algorithms. We do this until we have found an equilibrium
configuration with Haumea's mass. 
Outputs of the code include $\rho_{\rm core}$ and $\Omega$, and in general $\Omega$ will be 
smaller than Haumea's true angular frequency $\Omega_{\rm H} = 2\pi / P_{\rm rot}$ for $P_{\rm cmb} = 0$ MPa
($P_{\rm rot}$ is the rotational period).
We then repeat the procedure, finding an equilibrium configuration with Haumea's mass, having 
$P_{\rm cmb} = 40$ MPa. 
An output of the code will be a different $\rho_{\rm core}$ and a different $\Omega$, which in
general will be $> \Omega_{\rm H}$. 
If the solution is bracketed, we use standard bisection techniques to find $P_{\rm cmb}$ that
yields an equilibrium configuration consistent not just with Haumea's mass, but with its period 
as well. 

Thus, inputs of the code include $r_A = a$ and $r_B = b$, and Haumea's mass $M_{\rm H}$ and angular 
velocity $\Omega_{\rm H}$. 
The outputs of the code include the density $\rho$ at all locations, the calculated mass $M$
(which should comply with $M = M_{\rm H}$), the angular frequency $\Omega$ (which should equal
$2\pi / P_{\rm rot}$), and the values of $c$ and $P_{\rm cmb}$ and $\rho_{\rm core}$.

To benchmark the {\tt kyushu} code, we ensure that it reproduces a Jacobi ellipsoid when a homogeneous density is assumed.
A Jacobi ellipsoid with Haumea's mass and rotation period, and an axis $a = 960$ km, would have axes 
$b = 774.2$ km, $c = 498.8$ km, and uniform density $2579.7 \, {\rm kg} \, {\rm m}^{-3}$: this is similar to, but does not exactly equal, the solution favored by \citet{Lockwood2014}, who fit
Haumea's light curve assuming $a = 960$ km, $b = 770$ km, $c = 495$ km, and uniform density of $2600 \, {\rm kg} \, {\rm m}^{-3}$.
Running {\tt kyushu} and assuming axes of $a = 960$ km and $b = 774.2$ km
($b / a = 0.806$), the code finds an acceptable solution after about 7 bracketing iterations.
The mass matches Haumea's mass to within $0.03\%$ and the rotation period to within $0.04\%$.
The solution found is one with a very low value of $P_{\rm cmb} = 0.42 \, {\rm MPa}$, so that the body is essentially 
uniform in density, with an ice layer $< 5$ km in thickness (the resolution of the code).
The interior of the body has uniform density $2580.4 \, {\rm kg} \, {\rm m}^{-3}$, and the short axis has length
$c = 499.5$ km ($c / a = 0.520$). 
The density and $c$-axis match a uniform-density Jacobi ellipsoid to within $0.03\%$ and $0.14\%$, respectively. 
The code is therefore capable of finding the analytical solution of a uniform-density Jacobi ellipsoid, if the 
imposed $a$ and $b$ axes are consistent with such a solution.



\section{Results\label{sec:4}}

We have run the {\tt kyushu} code for 30 different combinations of $a$ and $b$ axes, with $a$
varying from 950 km to 1075 km in increments of $\sim$25 km, and $b$ varying from 800 km to 900 km in 
increments of $\sim$25 km. In comparing a subset of these runs with runs performed with 10 km increments, we found that convergence of key outputs (axes, densities) is 0.3\% or less, so we consider 25 km to be numerically converged.
We find families of solutions that can conform to Haumea's mass ($M = 4.006 \times 10^{21} \, {\rm kg}$)
and rotation period ($P_{\rm rot} = 3.9155 \, {\rm hr}$). 
Output quantities include the core density, the average (bulk) density, the outer $c$ axis, the shape of
the core-mantle boundary (CMB), and the thickness of the ice layer above the core. 

In Figure 2 we plot the following quantities as functions of imposed $a$ and $b$ axes:
the average (bulk) density, $\rho_{\rm avg}$; the (semi-)axis $c$; and the thickness of the ice layer 
along the $a$, $b$ and $c$ axes.
The density of the ice layer was imposed to be $921 \, {\rm kg} \, {\rm m}^{-3}$. 
In Figure 3 we plot as functions of $a$ and $b$ the following: 
the core density, $\rho_{\rm core}$; the pressure at the core-mantle boundary, $P_{\rm cmb}$; and 
the semi-axes $a_{\rm c}$, $b_{\rm c}$ and $c_{\rm c}$ of the core. 
Entire families of solutions are found across the range of $a$ and $b$ that we explored, with the exception of simultaneous combinations of large $a$ and large $b$. The input $a\times b$ combinations of $1025\times900,  1050\times900, 1075\times900$, and $1075\times800$ did not yield solutions because when initializing with these parameters, {\tt kyushu} was not able converge at Haumea's mass and rotation rate.
Simultaneously imposing large $a$ and $b$ yields a low bulk density $\rho_{\rm avg}$ and large parameter
$\omega^2 / (\pi G \rho_{\rm avg})$ that cannot yield a $c$ axis consistent with Haumea's mass.

\begin{figure*}
\centering
\includegraphics[width=0.95\textwidth]{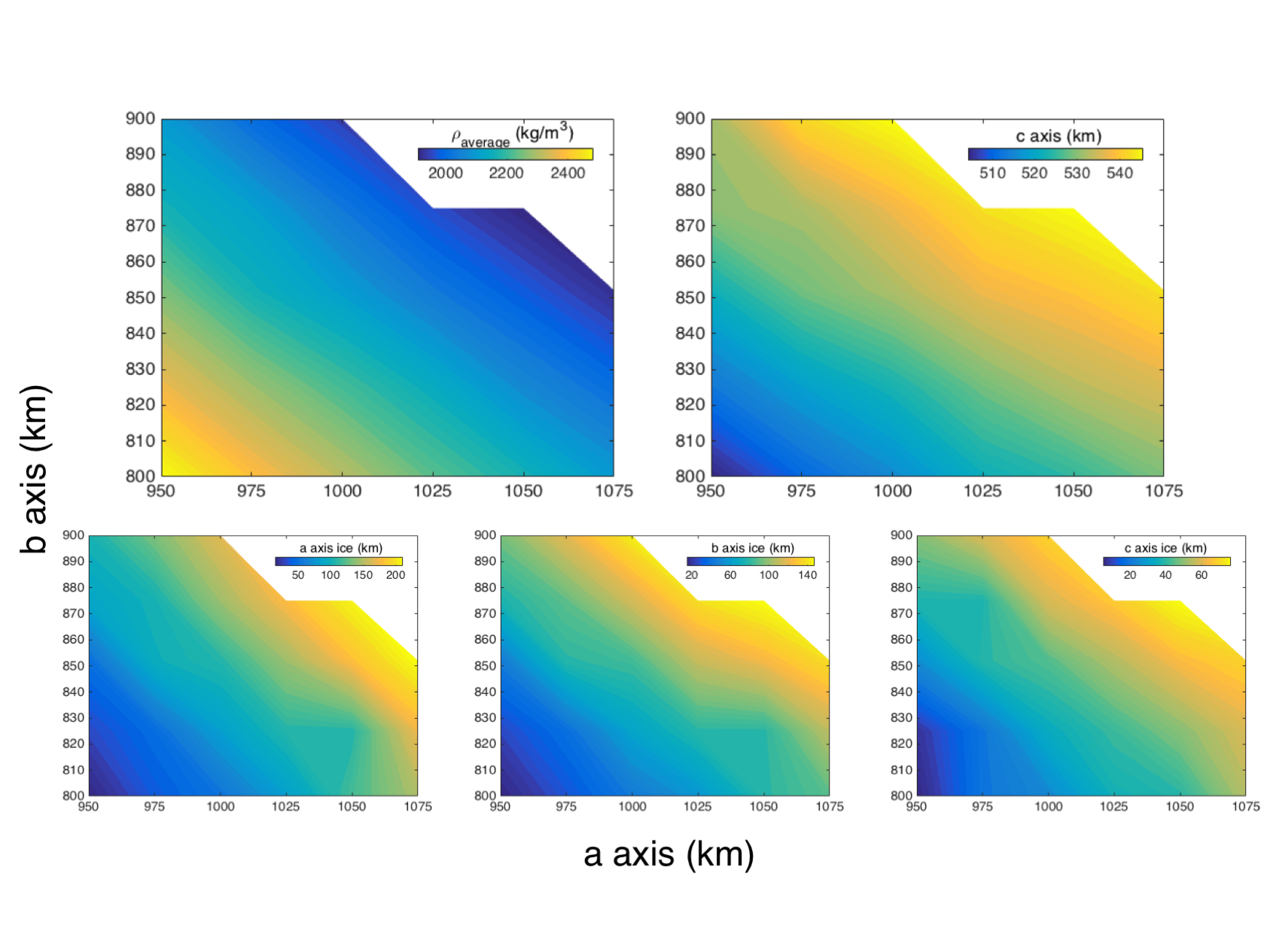}
\caption{Contour plots of quantities in the equilibrium shape models of a differentiated Haumea, as 
 functions of imposed axes $a$ and $b$.  The panels depict: the average (bulk) density of Haumea, $\rho_{\rm avg}$; the $c$ (semi-)axis; the thickness
 of the ice layer along the $a$ axis; the thickness of the ice layer along the $b$ axis; and the
 thickness of the ice layer along the $c$ axis. Solutions were not found for simultaneous combinations 
 of large $a$ and large $b$, but otherwise entire families of fluid hydrostatic equilibrium solutions exist.}
\label{figure:Hach_cont}
\end{figure*} 

\begin{figure*}
\centering
\includegraphics[width=0.95\textwidth]{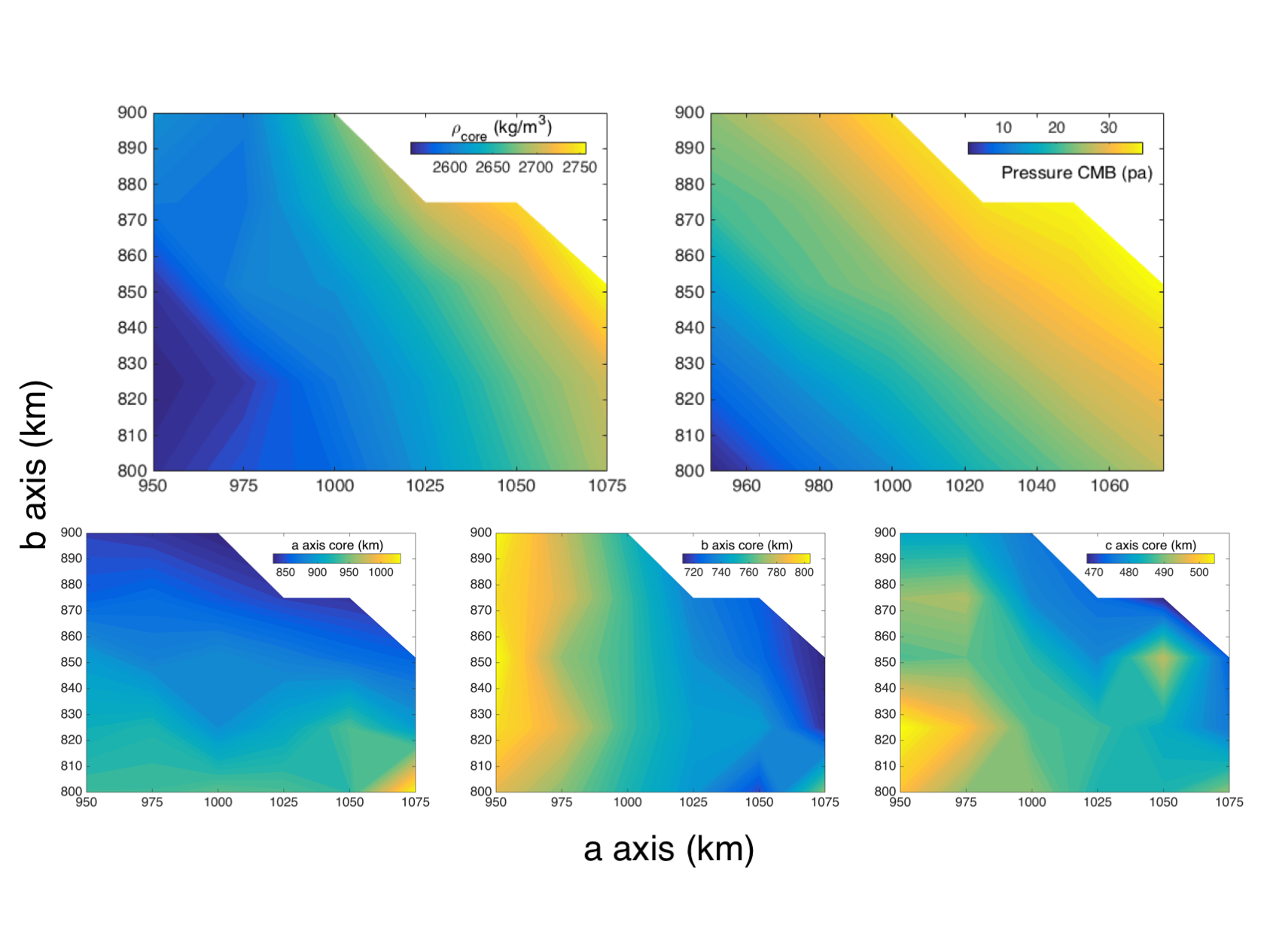}
\caption{Contour plots of quantities in the equilibrium shape models of a differentiated Haumea, as 
 functions of imposed axes $a$ and $b$.  The panels depict: the density of the core, $\rho_{\rm core}$; the pressure at the core-mantle boundary, $P_{\rm cmb}$, in MPa; the $a$ (semi-)axis of the core; the $b$ (semi-)axis of the core; and the $c$ (semi-)axis of the core. As in Figure 2, solutions were 
 not found for simultaneous combinations of large $a$ and large $b$, but otherwise entire families of fluid 
 hydrostatic equilibrium solutions exist.}
\label{figure:Hach_cont2}
\end{figure*} 

Across the explored parameter space that yielded solutions, the average density of Haumea ranges from 
1905 to $2495 \, {\rm kg} \, {\rm m}^{-3}$. As might be expected, the average (bulk) density of Haumea is equally sensitive to both $a$ and $b$, being inversely 
proportional to the volume and therefore to the product $a b$. As an example solution, we take $a = 1050$ km and $b = 840$ km, for which the average density is $2018 \, {\rm kg} \, {\rm m}^{-3}$.

For allowed solutions, the shortest (semi-)axis $c$ ranges from 504 to 546 km. 
The $c$ axis is equally sensitive to both $a$ and $b$, being large when the product $a b$ is large.
For the case with $a = 1050$ km and $b = 840$ km, we find $c = 537$ km. 

Likewise, the thickness of the ice layer increases with increasing $a$ and $b$ (and therefore $c$). 
The ice is always thickest along the $a$ axis, ranging from 15 to 210 km across the explored range;
it is intermediate in thickness along the $b$ axis, ranging from 15 to 150 km; 
and it is thinnest along the $c$ axis, ranging from 5 to 80 km.
For the case with $a = 1050$ km and $b = 840$ km, we find ice thicknesses of 167 km, 117 km, and 67 km along the
$a$, $b$, and $c$ axes. 

Across the explored parameter space that yielded solutions, the density of Haumea's core ranges from 
2560 to $2740 \, {\rm kg} \, {\rm m}^{-3}$.
Core density is much more sensitive to the $a$ axis than the $b$ axis, and tends to be greater when $a$ is greater. 
For the case with $a = 1050$ km and $b = 840$ km, we find a core density $2680\, {\rm kg} \, {\rm m}^{-3}$. 

Across the explored range, the pressure at the core mantle boundary ranges from 3.2 to 36.6 MPa, with the 
lowest values corresponding to the thinnest ice layers and smallest values of $a$ and $b$. 
For the case with $a = 1050$ km and $b = 840$ km, $P_{\rm cmb} = 30.4$ MPa. 
Finally, the size of the core ranges considerably. 
In general, the core approximates a triaxial ellipsoid, with the longest axis parallel to the $a$ axis.
For the extreme case with $a = 950$ km, $b = 800$ km, and $c = 504$ km (a small Haumea), 
we find core axes of $a_{\rm c} = 935$ km, $b_{\rm c} = 785$ km, and $c_{\rm c} = 499$ km. 
For this configuration, $b_{\rm c} / a_{\rm c} = 0.840$ and $c_{\rm c} / a_{\rm c} = 0.534$, very close 
to the axis ratios for the surface, $b / a = 0.842$ and $c / a = 0.531$. 
The mean size of the core, relative to the mean size of the surface, is $\xi = 0.991$.
That is, the ice layer thickness is only $\sim 1\%$ of the radius of Haumea, and comprises
$1.5\%$ of Haumea's volume.
For the opposite extreme case of a large Haumea, with $a = 1050$ km, $b = 875$ km, and $c = 546$ km, 
we find core axes of $a_{\rm c} = 840$ km, $b_{\rm c} = 725$ km, and $c_{\rm c} = 466$ km. 
For this configuration, $b_{\rm c} / a_{\rm c} = 0.86$ and $c_{\rm c} / a_{\rm c} = 0.55$, close 
to the axis ratios for the surface, $b / a = 0.83$ and $c / a = 0.52$. 
The mean size of the core, relative to the mean size of the surface, is $\xi = 0.888$,
meaning the ice layer thickness is $11\%$ the radius of Haumea and comprises $22\%$ of its volume. 
For the case we consider typical, with $a = 1050$ km and $b = 840$ km, the core semi-axes are $\approx 883 \times 723 \times 470$ km, which yields $b_{\rm c} / a_{\rm c} = 0.82$ and $c_{\rm c} / a_{\rm c} = 0.53$.
The mean size of the core, relative to the mean size of the surface, is $\xi = 0.909$, and the ice layer comprises $17.2\%$ of Haumea's volume.

In general, for larger assumed sizes of Haumea, the core becomes somewhat denser as Haumea itself becomes lower in density. The core takes up a smaller fraction of the volume of Haumea. While the core remains roughly similar in shape to the ellipsoid defined by the surface, there is a tendency for the core
to become slightly more spherical as Haumea's assumed size increases. 

To quantitatively test if Haumea's core and surface are both triaxial ellipsoids in our typical case, we calculated the maximum deviation from 1 of $(x/a_{c})^2 +(y/b_{c})^2+(z/c_{c})^2$, where $x$, $y$, and $z$ are computed for each angle combination $\theta$ and $\phi$ and radius defined by  $r=(\rm r(\rm ir_{\rm core}-1)+\rm r(\rm ir_{\rm core}))/2$. Here $ir_{\rm core}$ is the index of the first radial zone outside the core at that $\theta$ and $\phi$.  We find that a triaxial ellipsoid shape is consistent with the core to within 1.5\% and the surface to within 0.5\%. Both are within the code's resolution error. It is remarkable that the core mantle boundary solution is driven to the shape of a triaxial ellipsoid. This justifies the assumption in \S \ref{sec:2} that the surface and core would be triaxial ellipsoids if Haumea is differentiated.

\section{Discussion\label{sec:5}}

\subsection{Reconciling the light curve and occultation datasets}

Under the assumption of uniform density, there is only one Jacobi ellipsoid solution that can match Haumea's mass
and rotation rate, and its inferred $b / a$ axis ratio. 
Once $M$, $\omega$ and $b / a$ are specified, the average density $\rho_{\rm avg}$ is fixed, which determines $c / a$ and all the axes. 
In contrast, our modeling demonstrates that once the assumption of uniform density is dropped, a wide range of solutions exists 
with different semi-axes $a$ and $b$, and even with the same $b / a$ axis ratios. 
These would make linear cuts from the lower left to the upper right through the contour plots of Figures 2 and 3,
and would yield a range of $\rho_{\rm avg}$ and other properties.
This additional freedom suggests it may be possible to have a differentiated Haumea be a fluid in hydrostatic 
equilibrium, and simultaneously fit the shadow observed by \citet{Ortiz2017} during the occultation.

We find one solution, the example case considered above, to be quite favorable. 
This solution has outer semi-axes $a = 1050$ km, $b = 840$ km, and $c = 537$ km.
The core-mantle boundary is defined to lie at $P_{\rm c} = 30.4 \, {\rm MPa}$, and this surface is well approximated
by a triaxial ellipsoid with semi-axes $a_{\rm c} = 883$ km, $b_{\rm c} = 723$ km, and $c_{\rm c} = 470$ km.
The core density is $\rho_{\rm core} = 2680 \, {\rm kg} \, {\rm m}^{-3}$, and the average density of Haumea 
in this case is $\rho_{\rm avg} = 2018 \, {\rm kg} \, {\rm m}^{-3}$. 
The ice mantle in this case comprises 17.2\% of Haumea's volume and ranges in thickness
from 170 km on the $a$ axis, to 120 km on the $b$ axis, to 71 km on the $c$ axis.
The albedo is $p_{\rm V} \approx 0.66$, slightly lower than the range $0.71 - 0.84$ estimated by previous
studies, but higher than the value $\approx 0.51$ calculated by \citet{Ortiz2017}.
Likewise, the axes and average density we favor are intermediate between the previous solutions assuming a Jacobi 
ellipsoid with $\rho_{\rm avg} = 2600 \, {\rm kg} \, {\rm m}^{-3}$, and the \citet{Ortiz2017} case with 
$\rho_{\rm avg} = 1885 \, {\rm kg} \, {\rm m}^{-3}$.

The projection of this triaxial ellipsoid onto the Earth (its shadow) is a complicated function that depends
on its orientation relative to the line of sight. 
It is more difficult to invert the shadow axes to find the axes of Haumea's surface, but it is possible. 
We outline our methods in Appendix A (\ref{app}).

If the tilt of Haumea's rotation axis out of the plane of the sky is $\iota = 13.8^{\circ}$ and the rotational
phase were $\psi = 0^{\circ}$, we concur with \citet{Ortiz2017} that only a triaxial ellipsoid with axes 
$1161 \times 852 \times 513$ km would be consistent with shadow axes $852 \times 569$ km.
However, small changes in Haumea's rotational phase have a large impact on the shadow size. 
We find that if Haumea's rotational phase during the occultation were $\psi = 13.3^{\circ}$, then the shadow 
axes would be $a' = 853.1$ km and $b' = 576.8$ km, consistent with the observations by \citet{Ortiz2017} of  
$a' = 852 \pm 2$ km and $b' = 569 \pm 13$ km. 
\citet{Ortiz2017} favored $\psi = 0^{\circ}$, but inspection of their light curve (their Extended Data Figure 6)
shows that the rotational phase at minimum brightness was at least 0.04 ($14.4^{\circ}$), and would not be 
inconsistent with a value of 0.06 ($21.6^{\circ}$), relative to the phase of 0.00 at the time of the occultation. 
Finally we note that the axis ratio $b / a = 0.80$ for this case is the same as previous estimates 
($b / a \approx 0.80$; \citet{Lockwood2014}), and yields a light curve with $\Delta m \approx 0.23$ during the epoch of the occultation (\ref{app}). 
This approximates the actual light curve amplitude of $\Delta m = 0.26$ observed by \citet{Lockwood2014}. 

A much more extensive parameter study must be undertaken to simultaneously fit all the data. Astrometry of Haumea's moons can better constrain the moons' orbital poles and, if Hi'iaka's orbit is aligned
with Haumea's equator, Haumea's rotational pole and $\iota$. 
More and consistent analyses of the now 15 years of light curve data, especially considering different 
reflectance functions, can better constrain the rotational phase $\psi$ during the occultation. 
Further exploration of parameter space may yield a shape for Haumea that is exactly consistent with the
light curve data and the occultation shadow.
As shown here, though, Haumea can be a fluid in hydrostatic equilibrium and can conform to the occultation shadow.

\subsection{Aqueous alteration of Haumea's core and its astrobiological potential}

A large range of axes $a$ (from 950 to 1075 km) and $b$ (from 800 to 900 km) are consistent with bodies
with Haumea's mass and rotation rate, and are fluid configurations in hydrostatic equilibrium.
Our favored solution with $a = 1050$ km and $b = 840$ km has a mass fraction of ice of 17.2\%, but this value 
could range from 1\% to 22\% across the range we explored. 
Across this range, however, the allowable core density varies only slightly, from 
$\rho_{\rm core} = 2560 \, {\rm kg} \, {\rm m}^{-3}$ to $2740 \, {\rm kg} \, {\rm m}^{-3}$, deviating
by only a few percent from our favored density of $\rho_{\rm core} = 2680 \, {\rm kg} \, {\rm m}^{-3}$.
This is very close to the average density previously inferred for Haumea, but this appears to be coincidental. 
A robust result of our analysis is that Haumea's core has a density $\approx 2600 \, {\rm kg} \, {\rm m}^{-3}$, overlain by an ice mantle.

Comparison of the density of Haumea's core to other planetary materials provides strong clues to Haumea's history. Grain densities of ordinary and enstatite chondrites are typically $> 3600 \, {\rm kg} \, {\rm m}^{-3}$, and their 
bulk densities typically $\approx 3300 \, {\rm kg} \, {\rm m}^{-3}$ because of $\sim 10\%$ porosity 
\citep{Consolmagno2008,WILKISON2003}. 
Carbonaceous chondrites are marked by lower grain densities, average $ 3400 \, {\rm kg} \, {\rm m}^{-3}$ (range from $2400-5700 \, {\rm kg} \, {\rm m}^{-3}$ depending on the type of chondrite), higher porosities
$\approx 15-35\%$, and bulk densities closer to $2000 \, {\rm kg} \, {\rm m}^{-3}$ \citep{Macke2011,Consolmagno2008}. 
The difference is that carbonaceous chondrites are largely composed of products of aqueous alteration. In fact, the more oxidized groups of  carbonaceous chondrites have higher porosity \citep{Macke2011}.
Hydrated silicates typically have densities in this range.
Clays such as montmorillonite, kaolinite, illite, and mica typically have densities between 
$2600 \, {\rm kg} \, {\rm m}^{-3}$ and $2940 \, {\rm kg} \, {\rm m}^{-3}$ \citep{Osipov2012DENSITYMINERALS}. 
This strongly suggests that Haumea's core is composed of hydrated silicates, and that Haumea's core was aqueously
altered in its past. 

Serpentinization is the process by which silicates typical of the dust in the solar nebula react with water 
on an asteroid or planet, producing new phyllosilicate minerals.
A archetypal reaction would be:

\[
1.000 \, ({\rm Mg}_{0.71}{\rm Fe}_{0.29})_{2}{\rm SiO}_{4} + 1.142 \, {\rm H}_{2}{\rm O} \rightarrow
\]
\[
0.474 \, {\rm Mg}_{3}{\rm Si}_{2}{\rm O}_{5}({\rm OH})_{4} + 0.193 \, {\rm Fe}_{3}{\rm O}_{4} 
\]
\[
 + 0.052 \, {\rm SiO}_{2} + 0.194 \, {\rm H}_{2}.
\]
In this reaction, 1 kg of olivine with fayalite content typical of carbonaceous chondrites may react with 
0.129 kg of water to produce 0.826 kg of chrysotile, 0.281 kg of magnetite, 0.020 kg of silica, and 
0.002 kg of hydrogen gas, which escapes the system. 
The total density of olivine (density $3589 \, {\rm kg} \, {\rm m}^{-3}$) plus ice (density $ 921 \, {\rm kg} \, {\rm m}^{-3}$) 
before the reaction is $2697 \, {\rm kg} \, {\rm m}^{-3}$.
After the reaction, the mixture of chrysotile (density $2503 \, {\rm kg} \, {\rm m}^{-3}$) plus
magnetite (density $5150 \, {\rm kg} \, {\rm m}^{-3}$) plus silica (density $2620 \, {\rm kg} \, {\rm m}^{-3}$) 
has a total density of $2874 \, {\rm kg} \, {\rm m}^{-3}$ \citep{Coleman1971}.
Including a 10\% porosity typical of carbonaceous chondrites, the density of the aqueously altered system would be 
$2612 \, {\rm kg} \, {\rm m}^{-3}$, remarkably close to the inferred density of Haumea's core. 
%
%
%

If Haumea's core underwent significant aqueous alteration, some of this material may have dissolved in the water 
and ultimately found its way into the ice mantle of Haumea.
In fits to Haumea's reflectance spectrum, \citet{Pinilla-Alonso2009} found the most probable surface composition
was an intimate mixture of half crystalline and half amorphous water ice, with other components comprising $< 10\%$ 
of the surface; but similar modeling by \citet{Trujillo2007} found that Haumea's surface is best fit by a mixture
of roughly 81\% crystalline water ice and 19\% kaolinite.
Kaolinite was added to the fit to provide a spectrally neutral but blueish absorber; few other planetary
materials contribute to the reflectance spectrum in this way.
Kaolinite is a common clay mineral [${\rm Al}_{2}{\rm Si}_{2}{\rm O}_{5}({\rm OH})_{4}$] very similar in structure to chrysotile, produced by weathering of aluminum silicate minerals like feldspars. 

A variety of phyllosilicates have been observed by the Dawn mission on the surface of Ceres \citep{ammannito2016}, 
strongly suggesting aqueous alteration  of silicates within a porous, permeable core or a convecting mudball \citep{Bland2006, Travis2017OnCeres}.
If kaolinite can be confirmed in Haumea's mantle, this would provide strong support for the aqueous alteration of Haumea's core. 

Preliminary modeling by \citet{Neveu2015} suggests that aqueous alteration of Haumea's core is a very likely outcome.
Haumea, or its pre-collision progenitor, could have been differentiated into a rocky olivine core
and icy mantle. \citet{Neveu2015} show that many factors can lead to cracking of a rocky core on small bodies. 
Microcracking by thermal expansion mismatch of mineral grains or by thermal expansion of pore water during heating (as the core heats by radioactive decay over the first $< 1$ Gyr), will almost certainly introduce microfractures.
These would be widened by chemical reactions and water pressurization, etc., leading to macrofractures.
Cracks can heal by ductile flow of rock, but the rate is highly sensitive to temperature; below about 
750 K, healing of cracks takes longer than the age of the Solar System. 
Therefore it is highly likely that hydrothermal circulation of water through a cracked core would ensue.
Thermal modeling by \citet{Neveu2015} suggests Haumea's interior could be effectively fully convective,
allowing water and olivine to fully react and produce phyllosilicates. 
Circulation of water also would help cool the core, preventing temperatures from exceeding 750 K, ensuring
that fractures remain open, and that the hydrated silicates would not dehydrate. 
Liquid water is predicted to have existed for $\sim 10^8$ yr, although further geochemical modeling is needed to test more proposed scenarios for Haumea's structure and evolution.

A long ($\sim 10^8$ yr) duration of aqueous alteration suggests a period of habitability within Haumea.
To develop and survive, life as we know it requires water and a long-lasting environment with little temperature variability \citep{Davis1996}.
With central temperatures approaching 750 K, and surface temperatures near 40 K,  a large fraction of Haumea's interior would have had intermediate temperatures consistent with liquid water \citep{Rogez2012}. The origin of life is also thought to require a substrate to protect and localize biochemical reactions.
Clays such as montmorillonite can act as this substrate because they can bind substantial water, and are soft and delaminate easily. Clays can also promote the assembly of RNA from nucleosides, and can stimulate micelles to form vesicles \citep{Travis2017OnCeres}. The interior of Haumea may have at one point resembled regions beneath the seafloor experiencing hydrothermal circulation.
These regions are conducive to life: \citet{Czaja2016Sulfur-oxidizingAfrica} discovered archaea anaerobically metabolizing ${\rm H}_{2}{\rm S}$ in such environments, and other studies have confirmed that microbes exist deep in fractures
of hot environments \citep{Jannasch1985}.  

\subsection{Implications for the mass of the collisional family} 

An ongoing mystery is why Haumea's collisional family contains so little ice. The total masses of Hi'iaka and Namaka, plus 2002 TX$_{300}$ and the other collisional family members, amount to about 2.4\% of Haumea's mass \citep{Vilenius2018}. This is much smaller than the amount of ice that has been presumed to have been ejected. As described in \S\ref{sec:2}, if Haumea really were a Jacobi ellipsoid with uniform density $\approx 2600 \, {\rm kg} \, {\rm m}^{-3}$, it would have to have a very thin ice layer comprising perhaps only $\approx 4\%$ of Haumea's mass. 
This is much lower than the mass fraction of ice in typical KBOs. If the KBO has bulk density $\rho_0$, the mass fraction of ice would be $f_{\rm ice} = (\rho_{\rm ice})/(\rho_{0})\times(\rho_{\rm rock}-\rho_{0}) / (\rho_{\rm rock}-\rho_{\rm ice})$. A typical KBO may form from a mixture of pure olivine with 10\% porosity and density $\rho_{\rm rock} = 3300 \, {\rm kg} \, {\rm m}^{-3}$, and non-porous ice with density $\rho_{\rm ice} = 921 \, {\rm kg} \, {\rm m}^{-3}$.  The $\rho_0$ in such a KBO could range from $1500 \, {\rm kg} \, {\rm m}^{-3}$ to $2500 \, {\rm kg} \, {\rm m}^{-3}$ which equates to $f_{\rm ice}$ ranging from 46\% to 12\%.  If Haumea was comparable to these end member cases, it would need to lose 91\% and 67\% of its ice respectively to end up with a post-collisional ice fraction of 4\%. It is difficult to explain why Haumea would lose 91\% of its ice instead of 100\%. Also, neither of these scenarios match with the 2.4\% of ice thought to be ejected, which is also difficult to reconcile.

This discrepancy is ameliorated by our results. 
Our modeling of Haumea's structure shows that it may retained a significant fraction of ice. Across the parameter space we explored, Haumea's present-day bulk density varies from $1900 \, {\rm kg} \, {\rm m}^{-3}$ to $2500 \, {\rm kg} \, {\rm m}^{-3}$ (core density $2550 \, {\rm kg} \, {\rm m}^{-3}$ to $2750 \, {\rm kg} \, {\rm m}^{-3}$), which corresponds to $f_{\rm ice}$ ranging from 1\% to 22\%. The lower end of this range is unlikely from the standpoint of the occultation data. We favor that today, Haumea has a high ice fraction: $f_{\rm ice}$=17\% is our favored case.

In addition to this argument, our model suggests that Haumea underwent serpentinization, meaning the core experienced pervasive aqueous alteration. This process would reduce the fraction of ice below that which Haumea started. As an example, 
if Haumea initially had a density $\rho_0 = 2500 \, {\rm kg} \, {\rm m}^{-3}$, like that of Eris \citep{Brown2007a}, and original $\rho_{rock} = 3300 \, {\rm kg} \, {\rm m}^{-3}$, it started with $f_{\rm ice}$=34\%. Serpentinization would have then consumed ice into the rocky core to lower the core density to $\rho_{rock} = 2612 \, {\rm kg} \, {\rm m}^{-3}$, which would alter Haumea's ice fraction to 23\%. So, if the collision ejected 2.4\% of the ice, Haumea's ice fraction today would be $f_{\rm ice}\sim 20\%$. This estimate is within the range of ice fractions we predict from our parameter study. 

In conclusion, our modeling suggests both that Haumea may today retain a significant fraction of its original ice, and that some of the ice may have been lost to serpentinization of the core. Both of these factors imply that less ice needs to have been ejected for Haumea to have its present-day, observed ice fraction, possibly explaining the low total mass of the collisional family. 

\section{Conclusions}
This paper presents numerical modeling designed to test three questions about the KBO Haumea: 1) Is Haumea a Jacobi ellipsoid? If it is differentiated, what is Haumea's shape? 2) Is Haumea a fluid in hydrostatic equilibrium? 3) Can Haumea's occultation and light curve data be reconciled? We aimed to address these questions with the goal of understanding the composition and structure of Haumea to learn about its collisional history and evolution. 

We have written a code {\tt kyushu} based on the algorithms of Hachisu (H86a,b) to calculate the internal structure of a rapidly rotating differentiated body based on input parameters such as the semi-axes $a$ and $b$. Although we did not explore all parameter space, Haumea appears to be best approximated as a differentiated triaxial ellipsoid body in hydrostatic equilibrium with axes $a$ = 1050 km, $b$ = 840 km, and $c$ = 537 km. This shape fits the \citet{Ortiz2017} occultation shadow and is close to light curve data. As this shape, Haumea has core axes $a_{\rm c} = 883$ km, $b_{\rm c} = 723$ km, $c_{\rm c}=470$ km, $\rho_{\rm avg} = 2018 \, {\rm kg} \, {\rm m}^{-3}$, $\rho_{\rm core} = 2680 \, {\rm kg} \, {\rm m}^{-3}$ which equates to an ice mantle comprising $\sim 17\%$ of Haumea's mass and ranging from 71 to 170 km in thickness. Haumea's albedo is $p_{\rm v}\sim 0.66$ in this case. 

In contrast to previous studies \citep{Lockwood2014, Rabinowitz2006}, our results suggest that Haumea's ice crust amounts to a significant portion of the body. Due to the thickness of the ice, Haumea's core has a relatively high density indicating the composition of the core is a hydrated silicate (the closest match is kaolinite). For the core to be hydrated, a long period ($\sim 10^8$ yr) of serpentinization must have occured during which regions of the core were potentially habitable. The thick ice crust also suggests that Haumea's collisional family (icy objects a few percent the mass of Haumea) was produced from only a small portion of the ice Haumea started with, before Haumea suffered the collision. Insights into this type of mantle stripping collision could be applicable to modeling metal-rich, fast-rotating triaxial ellipsoid 16 Psyche, the focus of the upcoming NASA Psyche mission \citep{Elkins-Tanton2016}.

As this study continues, we would like to expand parameter space to obtain more precise results. We can explore how Haumea would change shape or composition if we use different ice densities, porosity, angles/orientations to better match the shadow in addition to matching the light curve amplitude/phase more precisely and using an appropriate equation of state to include the compressibility of materials. Haumea is a unique and interesting body worthy of study for its own sake, but understanding Haumea can provide insights into fundamental processes such as subsurface oceans/aqueous alteration on small bodies and dynamics of mantle-stripping collisions, acting across the Solar System.


\acknowledgments
 We thank Darin Ragozzine and Sarah Sonnett for helpful discussions about the collisional family and Haumea's light curve. 
We thank Steve Schwartz and Viranga Perera for useful discussions about how to model Haumea using smoothed particle hydrodynamics codes. We thank Leslie Rogers and Ellen Price for introducing us to the Hachisu (H86a,b) algorithm and for general discussions about how they implemented the Hachisu algorithm for exoplanets. We gratefully acknowledge partial support by the NASA Solar Systems Workings Program. 

\bibliography{Haumea.bib}

\begin{thebibliography}{}
\expandafter\ifx\csname natexlab\endcsname\relax\def\natexlab#1{#1}\fi

\bibitem[{Ammannito {et~al.}(2016)Ammannito, Desanctis, Ciarniello, Frigeri,
  Carrozzo, Combe, Ehlmann, Marchi, McSween, Raponi, Toplis, Tosi,
  Castillo-Rogez, Capaccioni, Capria, Fonte, Giardino, Jaumann, Longobardo,
  Joy, Magni, McCord, McFadden, Palomba, Pieters, Polanskey, Rayman, Raymond,
  Schenk, Zambon, \& Russell}]{ammannito2016}
Ammannito, E., Desanctis, M.~C., Ciarniello, M., {et~al.} 2016, Science, 353

\bibitem[{Binzel {et~al.}(1989)Binzel, Gehrels, \&
  Matthews}]{Binzel1989AsteroidsII}
Binzel, R.~P., Gehrels, T., \& Matthews, M.~S. 1989, {Asteroids II} (University
  of Arizona Press), 1258

\bibitem[{Birch(1947)}]{Birch2947}
Birch, F. 1947, Physical Review, 71, 809

\bibitem[{Bland {et~al.}(2006)Bland, Zolensky, Benedix, \& Sephton}]{Bland2006}
Bland, P., Zolensky, M., Benedix, G., \& Sephton, M. 2006, Meteorites and the
  Early Solar System II, 853

\bibitem[{Brown {et~al.}(2006)Brown, Barkume, Blake, Schaller, Rabinowitz, Roe,
  \& Trujillo}]{Brown2006}
Brown, M.~E., Barkume, K.~M., Blake, G.~A., {et~al.} 2006, The Astronomical
  Journal, 133, 284

\bibitem[{Brown {et~al.}(2007)Brown, Barkume, Ragozzine, \&
  Schaller}]{Brown2007}
Brown, M.~E., Barkume, K.~M., Ragozzine, D., \& Schaller, E.~L. 2007, Nature,
  446, 294

\bibitem[{Brown \& Schaller(2007)}]{Brown2007a}
Brown, M.~E., \& Schaller, E.~L. 2007, Science, 316, 1585

\bibitem[{Capitani \& Stixrude(2012)}]{Capitani2012}
Capitani, G.~C., \& Stixrude, L. 2012, American Mineralogist, 97, 1177

\bibitem[{Castillo-Rogez \& Lunine(2012)}]{Rogez2012}
Castillo-Rogez, J.~C., \& Lunine, J. 2012, in Frontiers of Astrobiology, ed.
  C.~Impey, J.~Lunine, \& J.~Funes (Cambridge University Press), 201

\bibitem[{Coleman(1971)}]{Coleman1971}
Coleman, R.~G. 1971, Bulletin of the Geological Society of America, 82, 897

\bibitem[{Consolmagno {et~al.}(2008)Consolmagno, Britt, \&
  Macke}]{Consolmagno2008}
Consolmagno, G.~J., Britt, D.~T., \& Macke, R.~J. 2008, Chemie der Erde, 68, 1

\bibitem[{Czaja {et~al.}(2016)Czaja, Beukes, \&
  Osterhout}]{Czaja2016Sulfur-oxidizingAfrica}
Czaja, A.~D., Beukes, N.~J., \& Osterhout, J.~T. 2016, 44, 983

\bibitem[{Davis \& McKay(1996)}]{Davis1996}
Davis, W.~L., \& McKay, C.~P. 1996, Origins of life and evolution of the
  biosphere, 26, 61

\bibitem[{Desch \& Neveu(2015)}]{Neveu2015}
Desch, S., \& Neveu, M. 2015, Lunar and Planetary Science Conference, {\#}2082

\bibitem[{Desch {et~al.}(2009)Desch, Cook, Doggett, \& Porter}]{Desch2009}
Desch, S.~J., Cook, J.~C., Doggett, T., \& Porter, S.~B. 2009, Icarus, 202, 694

\bibitem[{Desch \& Turner(2015)}]{Desch2015}
Desch, S.~J., \& Turner, N.~J. 2015, The Astrophysical Journal, 811, 156

\bibitem[{Elkins-Tanton {et~al.}(2016)Elkins-Tanton, Asphaug, Bell, Bercovici,
  Bills, Binzel, \& Bottke}]{Elkins-Tanton2016}
Elkins-Tanton, L.~T., Asphaug, E., Bell, J., {et~al.} 2016, Lunar and Planetary
  Science Conference, {\#}1631

\bibitem[{Hachisu(1986{\natexlab{a}})}]{Hachisu1986a}
Hachisu, I. 1986{\natexlab{a}}, The Astrophysical Journal Supplement Series,
  61, 479

\bibitem[{Hachisu(1986{\natexlab{b}})}]{Hachisu1986}
---. 1986{\natexlab{b}}, The Astrophysical Journal Supplement Series, 62, 461

\bibitem[{Holsapple(2001)}]{Holsapple2001}
Holsapple, K.~A. 2001, Icarus, 154, 432

\bibitem[{Jannasch \& Mottl(1985)}]{Jannasch1985}
Jannasch, H., \& Mottl, M. 1985, Science, 717

\bibitem[{Lacerda \& Jewitt(2006)}]{Lacerda2006}
Lacerda, P., \& Jewitt, D. 2006, The Astronomical Journal, 133, 13

\bibitem[{Lacerda {et~al.}(2008)Lacerda, Jewitt, \& Peixinho}]{Lacerda2008}
Lacerda, P., Jewitt, D., \& Peixinho, N. 2008, The Astronomical Journal, 135,
  1749

\bibitem[{Lellouch {et~al.}(2010)Lellouch, Kiss, Santos-Sanz, M{\"{u}}ller,
  Fornasier, {Groussin}, Lacerda, Ortiz, Thirouin, {Delsanti}, Duffard, Harris,
  Henry, Lim, {Moreno}, Mommert, Mueller, Protopapa, Stansberry, Trilling,
  Vilenius, Barucci, {Crovisier}, {Doressoundiram}, Dotto, Guti{\'{e}}rrez,
  {Hainaut}, Hartogh, {Hestroffer}, {Horner}, {Jorda}, {Kidger}, {Lara},
  Rengel, Swinyard, \& {Thomas}}]{Lellouch2010}
Lellouch, E., Kiss, C., Santos-Sanz, P., {et~al.} 2010, Astronomy and
  Astrophysics, 518, L147

\bibitem[{Lockwood {et~al.}(2014)Lockwood, Brown, \& Stansberry}]{Lockwood2014}
Lockwood, A.~C., Brown, M.~E., \& Stansberry, J. 2014, Earth, Moon, and
  Planets, 111, 127

\bibitem[{Macke {et~al.}(2011)Macke, Consolmagno, \& Britt}]{Macke2011}
Macke, R.~J., Consolmagno, G.~J., \& Britt, D.~T. 2011, Meteoritics and
  Planetary Science, 46, 1842

\bibitem[{N{\'{u}}{\~{n}}ez-Valdez {et~al.}(2013)N{\'{u}}{\~{n}}ez-Valdez, Wu,
  Yu, \& Wentzcovitch}]{Nunez-Valdez2013}
N{\'{u}}{\~{n}}ez-Valdez, M., Wu, Z., Yu, Y.~G., \& Wentzcovitch, R.~M. 2013,
  Geophysical Research Letters, 40, 290

\bibitem[{Ortiz {et~al.}(2017)Ortiz, Santos-Sanz, Sicardy, Benedetti-Rossi,
  Berard, Morales, Duffard, Braga-Ribas, \& Hopp}]{Ortiz2017}
Ortiz, J.~L., Santos-Sanz, P., Sicardy, B., {et~al.} 2017, Nature Publishing
  Group, 550, 219

\bibitem[{Osipov(2012)}]{Osipov2012DENSITYMINERALS}
Osipov, V.~I. 2012, Soil Mechanics and Foundation Engineering, 48, 8

\bibitem[{Pinilla-Alonso {et~al.}(2009)Pinilla-Alonso, Brunetto, Licandro,
  Gil-Hutton, Roush, \& Strazzulla}]{Pinilla-Alonso2009}
Pinilla-Alonso, N., Brunetto, R., Licandro, J., {et~al.} 2009, Astronomy and
  Astrophysics, 496, 547

\bibitem[{Price \& Rogers(2019)}]{Price2019}
Price, E.~M., \& Rogers, L.~A. 2019, The Astrophysical Journal, Submitted

\bibitem[{Probst(2015)}]{Probst2015}
Probst, L. 2015, Arizona State University Masters Thesis

\bibitem[{Rabinowitz {et~al.}(2006)Rabinowitz, Barkume, Brown, Roe, Schwartz,
  Tourtellotte, \& Trujillo}]{Rabinowitz2006}
Rabinowitz, D.~L., Barkume, K.~M., Brown, M.~E., {et~al.} 2006, The
  Astrophysical Journal, 639, 1238

\bibitem[{Ragozzine \& Brown(2009)}]{Ragozzine2009}
Ragozzine, D., \& Brown, M.~E. 2009, The Astronomical Journal, 137, 4766

\bibitem[{Shaw(1986)}]{Shaw1986}
Shaw, G.~H. 1986, The Journal of Chemical Physics, 84, 5862

\bibitem[{Sheppard \& Jewitt(2002)}]{Sheppard2002}
Sheppard, S.~S., \& Jewitt, D.~C. 2002, The Astronomical Journal, 124, 1757

\bibitem[{Travis(2017)}]{Travis2017OnCeres}
Travis, B. 2017, Astrobiology Science Conference, {\#}3620

\bibitem[{Trujillo {et~al.}(2007)Trujillo, Brown, Barkume, Schaller, \&
  Rabinowitz}]{Trujillo2007}
Trujillo, C.~A., Brown, M.~E., Barkume, K.~M., Schaller, E.~L., \& Rabinowitz,
  D.~L. 2007, The Astrophysical Journal, 655, 1172

\bibitem[{Vilenius {et~al.}(2018)Vilenius, Stansberry, Muller, Mueller, Kiss,
  Santos-Sanz, Mommert, Pal, Lellouch, Ortiz, Peixinho, Thirouin, Lykawka,
  Horner, Duffard, Fornasier, \& Delsanti}]{Vilenius2018}
Vilenius, E., Stansberry, J., Muller, T., {et~al.} 2018, Astronomy {\&}
  Astrophysics, 136, 1

\bibitem[{Vinet {et~al.}(1987)Vinet, Ferrante, Rose, \& Smith}]{Vinet1987}
Vinet, P., Ferrante, J., Rose, J.~H., \& Smith, J.~R. 1987, Geophysical
  Research Letters, 92, 9319

\bibitem[{Volk \& Malhotra(2012)}]{Volk2012}
Volk, K., \& Malhotra, R. 2012, Icarus, 221, 106

\bibitem[{Wilkison {et~al.}(2003)Wilkison, McCoy, McCamant, Robinson, \&
  Britt}]{WILKISON2003}
Wilkison, S.~L., McCoy, T.~J., McCamant, J.~E., Robinson, M.~S., \& Britt, D.
  2003, Meteoritics {\&} Planetary Science, 38, 1533

\end{thebibliography}

\appendix
\label{app}

\newcommand{\nhat}{\mbox{$\hat{n}$}}
\newcommand{\rvec}{\mbox{\boldmath$r$}}
\newcommand{\exhat}{\mbox{$\hat{e}_{x}$}}
\newcommand{\eyhat}{\mbox{$\hat{e}_{y}$}}
\newcommand{\ezhat}{\mbox{$\hat{e}_{z}$}}
\newcommand{\eloshat}{\mbox{$\hat{e}_{\rm LOS}$}}
\newcommand{\eonehat}{\mbox{$\hat{e}_{1}$}}
\newcommand{\etwohat}{\mbox{$\hat{e}_{2}$}}

Here we derive the formulas needed to calculate the axes of Haumea's shadow as it occults a star.
We assume Haumea's surface is a triaxial ellipsoid with long axis along the $x$ direction, with axes $a > b > c$, 
defined by those points that satisfy 
\[
f(x,y,z) = \frac{x^2}{a^2} +\frac{y^2}{b^2} +\frac{z^2}{c^2} = 1.
\]
We assume the star lies in a direction 
\[
\eloshat = \cos \psi \, \sin \phi \, \exhat +\sin \psi \, \sin \phi \, \eyhat + \cos \phi \, \ezhat  
\]
Here $\phi$ is the angle between the line of sight (from us through Haumea to the star) and Haumea's pole (along the $z$ axis).
We can define two unit vectors in the plane of the sky: 
\[
\eonehat = -\sin \psi \, \exhat + \cos \psi \, \eyhat,
\]
and
\[
\etwohat = \eonehat \times \eloshat = +\cos \psi \cos \phi \, \exhat +\sin \psi \cos \phi \, \eyhat -\sin \phi \, \ezhat.
\]

Haumea's limb is the locus of those points, defined by $\rvec$, such that the line of sight is tangent to the surface, 
or perpendicular to the normal, so that 
\[
\nabla f \, \cdot \, \eloshat = 0. 
\]
All of these points satisfy 
\[
z = -c^2 \, \tan \phi \, \left( \frac{x \, \cos \psi}{a^2} + \frac{y \, \sin \psi}{b^2} \right),
\]
which define a plane inclined to the sky. 
The intersection of the plane with the ellipsoid defines an ellipse, and the projection of this ellipse onto the plane
of the sky---Haumea's shadow---also is an ellipse.

We project the points on Haumea's limb onto the plane of the sky by recasting $\rvec$ in the coordinate system using 
$\eonehat$, $\etwohat$, and $\eloshat$:
\[
\rvec = \left( \rvec \cdot \eonehat \right) \, \eonehat + \left( \rvec \cdot \etwohat \right) \, \etwohat
      + \left( \rvec \cdot \eloshat \right) \, \eloshat = s \, \eonehat + t \, \etwohat + u \, \eloshat, 
\]
with 
\[
s = \rvec \cdot \eonehat = -x \, \sin \psi + y \, \cos \psi
\]
and
\[
t = \rvec \cdot \etwohat = x \, \cos \psi \, \cos \phi + y \, \sin \psi \, \cos \pi - z \, \sin \phi.
\]
All the points on the limb have $z$ related to $x$ and $y$ as above, so the boundary of the shadow, which
equals the projection of the limb onto the plane of the sky, is defined by 
\[
s = -x \, \sin \psi + y \, \cos \psi
\]
and
\[
\frac{t}{\cos \phi} = x \, \cos \psi \, \left( 1 + \frac{c^2}{a^2} \, \tan^2 \phi \right) 
                     +y \, \sin \psi \, \left( 1 + \frac{c^2}{b^2} \, \tan^2 \phi \right).
\]

Inverting, we find $x$, $y$ and $z$ in terms of $s$ and $t$ for points along the limb:
\[
x = \frac{1}{\Delta} \, \left[ -\sin \psi \, \left( 1 + \frac{c^2}{b^2} \, \tan^2 \phi \right) \, s 
 + \cos \psi \, \frac{t}{\cos \phi} \right],
\]
\[
y = \frac{1}{\Delta} \, \left[ +\cos \psi \, \left( 1 + \frac{c^2}{a^2} \, \tan^2 \phi \right) \, s 
 + \sin \psi \, \frac{t}{\cos \phi} \right],
\]
and 
\[
z = -\frac{\tan \phi}{\Delta} \, \left[ \sin \psi \, \cos \psi \, \left( \frac{c^2}{b^2}-\frac{c^2}{a^2} \right)
 + \left( \frac{c^2}{a^2} \, \cos^2 \psi + \frac{c^2}{b^2} \, \sin^2 \psi \right) \, \left( \frac{t}{\cos \phi} \right) \right],
\]
where
\[
\Delta = 1 + c^2 \, \tan^2 \phi \, \left( \frac{ \cos^2 \psi }{ a^2 } + \frac{ \sin^2 \psi }{ b^2 } \right).
\]

Subsituting these expressions for $x$, $y$ and $z$ into the equation for the ellipsoidal surface, we find the projection
of Haumea's limb onto the plane of the sky satisfies 
\[
P s^2 + Q s t + R t^2 = 1,
\]
where 
\[
P = \frac{1}{\Delta^2 \, a^2} \, \sin^2 \psi \, \left( 1 + \frac{c^2}{b^2} \, \tan^2 \phi \right)^2
   +\frac{1}{\Delta^2 \, b^2} \, \cos^2 \psi \, \left( 1 + \frac{c^2}{a^2} \, \tan^2 \phi \right)^2
\]
\[
   +\frac{\tan^2 \phi}{\Delta^2 \, c^2} \, \sin^2 \psi \, \cos^2 \psi \, \left( \frac{c^2}{b^2} -\frac{c^2}{a^2} \right)^2,
\]
\[
R = \frac{1}{\Delta^2 \, \cos^2 \phi} \, \left( \frac{\cos^2 \psi}{a^2} + \frac{\sin^2 \psi}{b^2} \right) 
 + \frac{c^2 \, \tan^2 \phi}{\Delta^2 \, \cos^2 \phi} \, \left( \frac{\cos^2 \psi}{a^2} + \frac{\sin^2 \psi}{b^2} \right)^2,
\]
and 
\[
Q = \frac{2 \, \sin \psi \, \cos \psi}{\Delta^2 \, c^2 \, \cos \phi} \, \left( \frac{c^2}{b^2} -\frac{c^2}{a^2} \right) \,
 \left[ 1 + c^2 \, \tan^2 \phi \, \left( \frac{\cos^2 \psi}{a^2} + \frac{\sin^2 \psi}{b^2} \right) \right].
\]
These also define an ellipse, rotated in the $s$-$t$ plane.

After rotating the ellipse in the plane of the sky by an angle $\theta$, defined by 
\[ 
\tan 2\theta = Q / (R-P),
\]
we find it has axes $a'$ and $b'$ defined by 

\[
\frac{1}{(a')^2} = \left[ P \, \cos^2 \theta + R \, \sin^2 \theta -Q \, \sin \theta \, \cos \theta \right]
\]
and 
\[
\frac{1}{(b')^2} = \left[ P \, \sin^2 \theta + R \, \cos^2 \theta +Q \, \sin \theta \, \cos \theta \right].
\]
We have written a simple code that takes $a$, $b$, and $c$, and $\psi$ and $\phi$ as inputs, and solves for $\theta$ 
and then the semi-axes $a'$ and $b'$ of Haumea's shadow. 

One end-member case includes $\phi = 0^{\circ}$, in which Haumea's pole is pointed toward the star;
we derive $\theta = -\psi$ and regardless of $\psi$, Haumea's shadow has axes $a' = b$ and $b' = a$.
Another end-member case is $\phi = 90^{\circ}$, in which case the line of sight to the star is parallel to Haumea's equator.
The shadow will have $b' = c$ regardless of $\psi$, and the other axis will be 
\[
a' = a b \, \left[ \frac{\cos^2 \psi}{a^2} + \frac{\sin^2 \psi}{b^2} \right]^{1/2},
\]
in which case $a' = b$ if $\psi = 0^{\circ}$ (looking along the long $a$ axis), or $a' = a$ if $\psi = 90^{\circ}$ 
(looking along the $b$ axis). 
One more end-member case is $\psi = 0^{\circ}$ but arbitrary $\phi$, in which case $a' = b$ and 
\[
b' = a \, \cos \phi \, \left[ 1 + \frac{c^2}{a^2} \, \tan^2 \phi \right]^{+1/2}.
\]
This is the case considered by Ortiz et al.\ (2017).
Assuming $a = 1161$ km, $b = 852$ km, $c = 513$ km, $\psi = 0^{\circ}$ and $\phi = 76.3^{\circ}$ (a tilt of Haumea's pole 
with respect to the plane of the sky of $13.7^{\circ}$), we find 
$a' = 852$ km and $b' = 584$ km, similar to the solution found by \citet{Ortiz2017}.

\end{document}